# How does the elimination of group mean-differences affect factor score determinacy?

André Beauducel[*] & Norbert Hilger

*Department of Psychology, University of Bonn, Germany*

**Abstract**

The present study investigates to what degree the common variance of the factor score predictor with the original factor, i.e., the determinacy coefficient or the validity of the factor score predictor, depends on the mean-difference between groups. When mean-differences between groups in the factor score predictor are eliminated by means of covariance analysis, regression, or group specific norms, this may reduce the covariance of the factor score predictor with the common factor. It is shown that in a one-factor model with the same group mean-difference on all observed variables, the common factor cannot be distinguished from a common factor representing the group mean-difference. It is also shown that for common factor loadings equal or larger than .60, the elimination of a $d = .50$ mean-difference between two groups in the factor score predictor leads to only small decreases of the determinacy coefficient. A compensation-factor $k$ is proposed allowing for the estimation of the number of additional observed variables necessary to recover the size of the determinacy coefficient before elimination of a group mean-difference. It turns out that for factor loadings equal or larger than .60 only a few additional items are needed in order to recover the initial determinacy coefficient after the elimination of moderate or large group mean-differences.

[*] Address for correspondence: André Beauducel, Department of Psychology, University of Bonn, Kaiser-Karl-Ring 9, 53111 Bonn, Germany; email: beauducel@uni-bonn.de

Factor score predictors are computed in order to provide individual scores for a factor resulting from factor analysis (DiStefano, Zhu, & Mindrila, 2009). These scores can be used in applied settings, e.g., job-selection. The validity of factor score predictors depends on their correlation with the corresponding factors, which is usually termed determinacy coefficient (Gorsuch, 1983; Grice, 2001). For orthogonal factors and block-diagonal patterns of factor loadings, the reliability of the regression factor score predictor also corresponds to the determinacy coefficient (Beauducel, Harms, & Hilger, 2016). However, under more general conditions, the correlation of the factor score predictor with the corresponding factor is an indicator of validity.

Several group mean-differences may co-occur with the individual differences represented by a common factor. Some group mean-differences may be estimated *a priori* in the context of multiple-group factor analysis, which leads to the issue of measurement invariance of factors across groups. Measurement invariance refers to the similarity of model parameters across groups, an assumption required for the interpretation of group differences on the factors (Vandenberg, & Lance, 2000). New methods that may allow for the improvement of measurement invariance have meanwhile been proposed (Asparouhov, & Muthén, 2023). However, not all relevant group mean-differences may be considered right from the start when factor analysis is performed. A reason for this could be that sample size leads to a limitation of the number of groups that can be considered in multiple-group factor analysis. Another reason could be that a grouping of individuals is obtained afterwards (e.g., by means of latent class analysis). A third reason could be that group mean-differences that appear of irrelevant size in the observed variables, can be of relevant size in the factor score predictors. Group mean-differences that are not represented by a model parameter for the means, may affect the covariances of observed variables, and may be present in the factor score predictor. In these cases, an elimination of group mean-differences may be performed *a posteriori* by means of regression analysis, covariance analysis, or by means of group-specific test norms or test standardization (American Educational Research Association, American Psychological Association, & National Council on Measurement in Education, 2014). The elimination of the group mean-differences may affect the validity of the factor score predictor, i.e., the factor score determinacy. Therefore, the present study investigates the effect of eliminating group mean-differences on the resulting determinacy coefficient.

In principle, different types of factor score predictors, different numbers of groups, different factor models, and different score distributions might be of interest in this context. However, in order to provide initial results that could be the starting point for further investigations, the present study is based on only two groups and on the determinacy coefficient computed for the best-linear factor score predictor (Krijnen, Wansbeek, & Ten Berge, 1996), which is sometimes termed regression factor score predictor. Moreover, the study is based on one factor models with equal factor loadings. These simplifications allow for a direct investigation of effects of the size of group mean-differences in terms of different effect sizes $d$ (Cohen, 1988) and the number of variables on the determinacy coefficient.

First, a population factor model for the description of the effect of group mean-differences on factors and model parameters is defined. On this basis, the effect of factor loadings and the number of variables on the determinacy coefficient is investigated. Benchmarks for factor loadings combined with a given number of observed variables resulting in reasonable determinacy coefficients are provided. Using these benchmarks, the effect of group mean-differences and determinacy coefficients is analyzed. It is proposed to calculate the number of additional items necessary to compensate for reduced determinacy resulting from eliminating group mean-differences. Finally, theoretical and practical implications of the results are discussed.

## Definitions

In the population of individuals, the common factor model with an additional factor representing the means between two groups can be defined as

$$\mathbf{x} = \mathbf{\Lambda}_\xi \boldsymbol{\xi} + \mathbf{\Lambda}_g \mathbf{g} + \mathbf{\Psi} \boldsymbol{\varepsilon}, \tag{1}$$

where $\mathbf{x}$ is a standardized random vector of $p$ observed variables, $\boldsymbol{\xi}$ is a random vector of $q$ common factors, with $E(\boldsymbol{\xi}\boldsymbol{\xi}') = \mathbf{\Phi}$ and $diag(\mathbf{\Phi}) = \mathbf{I}_{q \times q}$ and $\mathbf{\Lambda}_\xi$ is the $p \times q$ matrix of common factor loadings. Moreover, $\mathbf{g}$ is a random vector representing the means of two groups, with $E(\mathbf{gg}') = 1$, $E(\mathbf{g}\boldsymbol{\xi}') = \mathbf{0}$, and $\mathbf{\Lambda}_g$ is the $p \times 1$ matrix of loadings of the observed variables on the factor representing the standardized group mean-difference $d$ (e.g., Cohen, 1988), with $\sigma_d = d/2$. As $E(\mathbf{gg}') = 1$ and $E(\mathbf{g}) = 0$ $\mathbf{g}$ corresponds to an effect-coded indicator variable. The loadings $\mathbf{\Lambda}_g$ represent the correlation of the observed variables with $\mathbf{g}$. Therefore, the effect size $d$ is converted into the correlation $r$ according to Ruscio (2008, Table 1). For groups of equal size, the conversion of $d$ into a correlation $r$ is $r = d(4+d^2)^{-1/2}$, so that the loadings are given by $\mathbf{\Lambda}_g = d(4+d^2)^{-1/2}\mathbf{1}$, where $\mathbf{1}$ is a $p \times 1$ unit-vector. For unequal group sizes $\mathbf{\Lambda}_g$ can be computed by $\mathbf{\Lambda}_g = d(p_1^{-1}p_2^{-1}+d^2)^{-1/2}\mathbf{1}$, where $p_1$ is the base rate of group 1 and $p_2$ is the base rate of group 2. The random vector $\boldsymbol{\varepsilon}$ represents $p$ unique factors, with $E(\boldsymbol{\varepsilon}\boldsymbol{\varepsilon}') = \mathbf{I}_{p \times p}$, $E(\boldsymbol{\varepsilon}\boldsymbol{\xi}') = \mathbf{0}$, $E(\boldsymbol{\varepsilon}\mathbf{g}') = \mathbf{0}$, and $\mathbf{\Psi}$ is a $p \times p$ diagonal, positive definite, unique factor loading matrix. This implies that

$$E(\mathbf{xx}') = \Sigma = \Lambda_\xi \Phi \Lambda_\xi' + \Lambda_\mathbf{g} \Lambda_\mathbf{g}' + \Psi^2 \qquad (2)$$

is a correlation matrix.

For $q = 1$ and when $d$ is the same for all observed variables, it is impossible to disentangle $\xi$ and $\mathbf{g}$, so that only a single common factor $\xi_\mathbf{g}$ combining $\xi$ and $\mathbf{g}$ can empirically be identified when the mean-difference between the groups occurs consistently across all observed variables. A model based on a single detectable common factor $\xi_\mathbf{g}$ comprising equal group mean-differences on the observed variables and individual differences can be written as

$$\mathbf{x} = \Lambda_{\xi\mathbf{g}} \xi_\mathbf{g} + \Psi \varepsilon, \text{ with } \Lambda_{\xi\mathbf{g}} = (\Lambda_\xi^2 + \Lambda_\mathbf{g}^2)^{1/2} = (\Lambda_\xi^2 + d^2(4+d^2)^{-1}\mathbf{1})^{1/2}. \qquad (3)$$

The determinacy coefficient is the correlation of a factor score predictor with the corresponding factor, i.e.,

$$\mathbf{P} = diag\left(E\left(diag\left(\mathbf{w'xx'w}\right)^{-1/2}\mathbf{w'x}\xi_\mathbf{g}'\right)\right) = diag\left(diag\left(\mathbf{w'\Sigma w}\right)^{-1/2}\mathbf{w'}\Lambda_{\xi\mathbf{g}}\right). \qquad (4)$$

Inserting $\Sigma^{-1}\Lambda_{\xi\mathbf{g}}\Phi$, the weights for the best-linear factor score predictor (Grice, 2001), for $\mathbf{w}$ in Equation 4 yields $\mathbf{P} = diag\left(\Phi\Lambda_{\xi\mathbf{g}}'\Sigma^{-1}\Lambda_{\xi\mathbf{g}}\Phi\right)^{1/2}$ and

$$\mathbf{P} = diag\left(\Lambda_{\xi\mathbf{g}}'\Sigma^{-1}\Lambda_{\xi\mathbf{g}}\right)^{1/2} \qquad (5)$$

for $q = 1$.

## Results

*Effect of factor loadings and p on the determinacy coefficient*

For $q = 1$ and all loadings $\Lambda_{\xi\mathbf{g}}$ being equal, it follows from $\mathbf{w} = \Sigma^{-1}\Lambda_{\xi\mathbf{g}}\Phi$ that all weights $\mathbf{w}$ are equal so that one can write $\mathbf{w} = w\mathbf{1}$, where $w$ is a scalar with $w \neq 0$ and $\mathbf{1}$ is a $p \times 1$ unit-vector, so that

$$\mathbf{P} = diag\left(w^2 \mathbf{1'\Sigma 1}\right)^{-1/2} w\mathbf{1'}\Lambda_{\xi\mathbf{g}} = diag\left(\mathbf{1'\Sigma 1}\right)^{-1/2} \mathbf{1'}\Lambda_{\xi\mathbf{g}}. \qquad (6)$$

When all weights are equal and greater zero and all loadings are equal and greater zero, and $q = 1$, it is possible to write $\Lambda_{\xi\mathbf{g}} = \lambda_{\xi\mathbf{g}}\mathbf{1}$, so that

$$\boldsymbol{\Sigma} = \begin{bmatrix} 1 & \lambda_{\xi g}^2 & \cdots & \lambda_{\xi g}^2 \\ \lambda_{\xi g}^2 & 1 & \ddots & \vdots \\ \vdots & \ddots & 1 & \lambda_{\xi g}^2 \\ \lambda_{\xi g}^2 & \cdots & \lambda_{\xi g}^2 & 1 \end{bmatrix}. \qquad (7)$$

As for $q = 1$ and equal factor loadings, the weights cancel out, Equation 6 can be written as

$$\mathbf{P} = \left(1 + (p-1)\lambda_{\xi g}^2\right)^{-1/2} \lambda_{\xi g} p^{1/2}, \qquad (8)$$

and the common variance of the scores with the corresponding factor is

$$\mathbf{P}^2 = \left(\lambda_{\xi g}^{-2} + p - 1\right)^{-1} p. \qquad (9)$$

*Benchmarks of factor loadings for reasonable determinacy*

Factor score predictors should at least have 50% of common variance with the corresponding factor. Entering $\lambda_{\xi g} = (p+1)^{-1/2}$ into Equation 9, yields

$$\mathbf{P}^2 = \left(\lambda_{\xi g}^{-2} + p - 1\right)^{-1} p = \left(2p\right)^{-1} p = 0.50. \qquad (10)$$

However, 80% or 90% of common variance of the factor score predictor with the corresponding factor will be a more reasonable benchmark. Entering $\lambda_{\xi g}^2 = 4(p+4)^{-1}$ into the slightly transformed Equation 10 yields

$$\mathbf{P}^2 = \left(\lambda_{\xi g}^{-2} + p - 1\right)^{-1} p = \left(\frac{5}{4} p\right)^{-1} p = 0.80, \qquad (11)$$

and entering $\lambda_{\xi g}^2 = 9(p+9)^{-1}$ yields $\mathbf{P}^2 = 0.90$. The intended loading size for a given $p$ and $\mathbf{P}^2$ can be computed by

$$\lambda_{\xi g}^2 = \left(\frac{p}{\mathbf{P}^2} - p + 1\right)^{-1}. \qquad (12)$$

For 90% of common variance of the factor score predictor with the factor and $p = 4$ a loading size of .69 is required, which is attainable in several areas (see Figure 1).

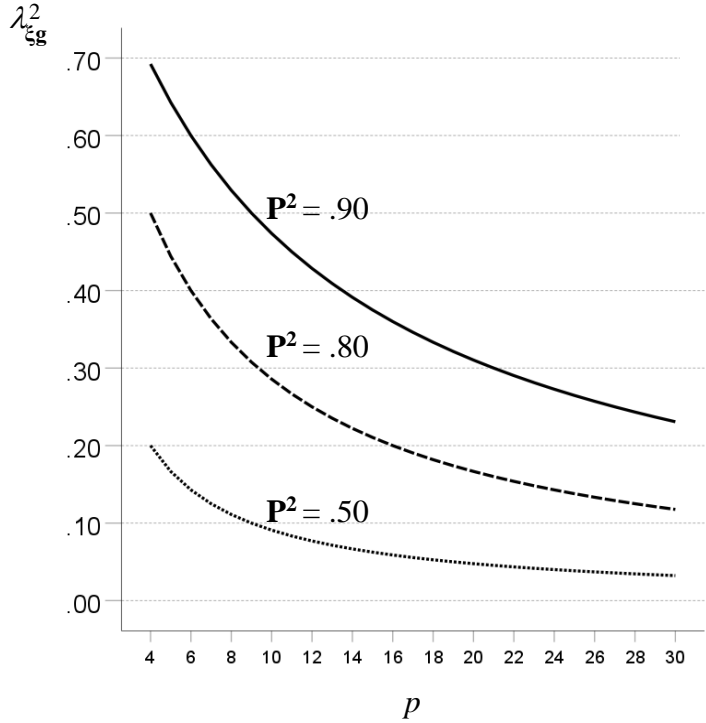

Figure 1. Loading size for different values of $p$ for 50%, 80%, and 90% of common variance of the factor score predictor with the factor

*Loading benchmark, group differences and determinacy*

In the following, the effect of eliminating group differences in the factor score predictor on the determinacy coefficient is considered for $q = 1$ and equal loading sizes. Eliminating group mean-differences implies that $\Lambda_g = d(4+d^2)^{-1/2}\mathbf{1} = \mathbf{0}$. It follows from Equation 3 that eliminating group mean-differences reduces $\Lambda_{\xi g}$ to $\Lambda_\xi$. It follows from $\Lambda_{\xi g}^2 = \Lambda_\xi^2 + d^2(4+d^2)^{-1}\mathbf{1}$ and for all loadings being equal in $\Lambda_{\xi g}$ that

$$\lambda_{\xi g}^2 = \lambda_\xi^2 + \frac{d^2}{4+d^2}. \tag{13}$$

Entering $\lambda_{\xi g}^2$ computed from Equation 13 into Equation 9 yields

$$\mathbf{P}^2 = \left( \frac{1}{\lambda_\xi^2 + \frac{d^2}{4+d^2}} + p - 1 \right)^{-1} p. \tag{14}$$

Entering $d = 0$ into Equation 14 yields the relationship between $\lambda_\xi$, $p$, and $\mathbf{P}^2$ as it is given in Equation 9 for $\lambda_{\xi g}$. Entering $\lambda_\xi = 0$ into Equation 14 yields the relationship between $d$, $p$, and $\mathbf{P}^2$ when there are no further inter-individual differences (see Figure 2 A). This condition is a baseline for Figure 2 B-F, where the relationship of *d* and *p* with $\mathbf{P}^2$ is presented for different levels of $\lambda_\xi$. Whereas the effect of eliminating a mean-difference of $d = 1.00$ on $\mathbf{P}^2$ is quite substantial for $\lambda_\xi = .20$ and $\lambda_\xi = .40$, it is rather small for $\lambda_\xi = .60$ and beyond (Figure 2 D-F).

However, according to Cohen (1988) $d = 0.80$ is a large effect, whereas $d = .50$ represents a moderate effect and $d = .20$ represents a small effect. Eliminating a moderate effect of $d = .50$ may also be considered. As an example, consider the condition $p = 10$ and $\lambda_\xi = .60$. For this condition and $d = 0.50$ Equation 14 yields

$$\mathbf{P}^2 = \left( \frac{1}{.60^2 + \frac{.50^2}{4+.50^2}} + 10 - 1 \right)^{-1} 10 = .88. \tag{15}$$

For $p = 10$, $\lambda_\xi = .60$, and $d = 0.00$ Equation 14 yields

$$\mathbf{P}^2 = \left( \frac{1}{.60^2} + 10 - 1 \right)^{-1} 10 = .85. \tag{16}$$

Thus, in this condition, eliminating a moderate effect size in the factor score predictor will reduce the common variance of the factor score predictor with the original factor by only 3%.

It might be possible to compensate a decrease of $\mathbf{P}^2$ resulting from eliminating a group mean-difference by means of an increase of *p*. The increase of *p* that necessary for compensating for the elimination of a group mean-difference can be calculated by means of the compensation-factor *k* as a multiplier of *p*. The compensation-factor *k* can be found by means of equating Equation 14 comprising the effects of d with Equation 14 for $d = 0$ and *k* as a multiplier of *p*. This yields

$$\left( \frac{1}{\lambda_\xi^2 + \frac{d^2}{4+d^2}} + p - 1 \right)^{-1} p = \left( \frac{1}{\lambda_\xi^2} + pk - 1 \right)^{-1} pk \tag{17}$$

and after some transformation

$$k = \left( \frac{1}{\lambda_\xi^2} - 1 \right) \left( \frac{1}{\lambda_\xi^2 + \frac{d^2}{4+d^2}} - 1 \right)^{-1}. \tag{18}$$

(A) $\lambda_\xi = .20$

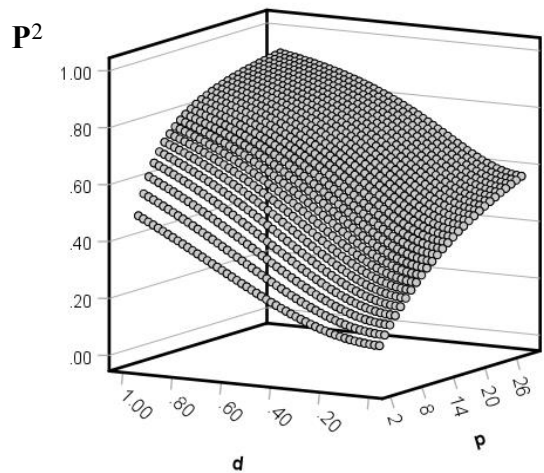

(B) $\lambda_\xi = .40$

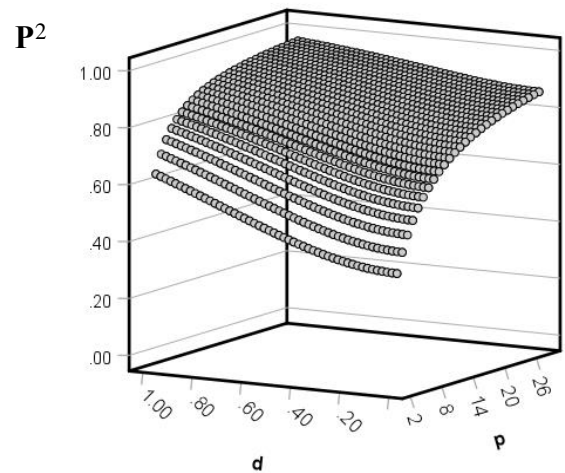

(C) $\lambda_\xi = .60$

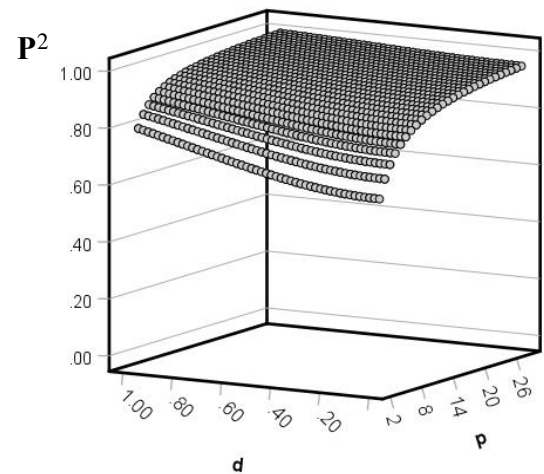

(D) $\lambda_\xi = .80$

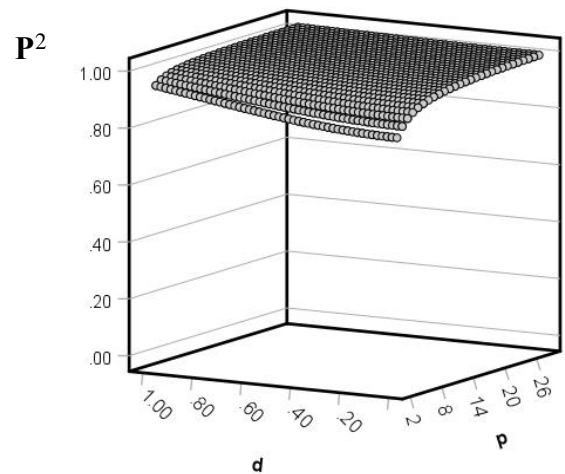

Figure 2. Effect of $0 \leq d \leq 1$ and $2 \leq p \leq 30$ on $\mathbf{P}^2$, for (A) $\lambda_\xi = .20$, (B) $\lambda_\xi = .40$, (C) $\lambda_\xi = .60$, and (D) $\lambda_\xi = .80$.

The difference $p^+ = kp - p$ might be rounded to the nearest integer, which indicates the number of observed variables that might be added in order to compensate for a decrease of $\mathbf{P}^2$ resulting from eliminating $d$. If $kp - p \geq 1$ it might be reasonable to add observed variables in order to reach the same $\mathbf{P}^2$ that occurred before elimination of the group mean-difference.

As an example, consider a factor score predictor based on $p = 5$, $\lambda_\xi = .60$, and $d = .81$, this results in $\mathbf{P}^2 = .83$ according to Equation 14. Now, $d$ is eliminated by means of separate test-norms, or something similar, so that $\mathbf{P}^2 = .74$. When $p = 5$ and $\lambda_\xi = .60$ are entered into Equation 18, the resulting $k$ is

$$k = \left(\frac{1}{.60^2} - 1\right)\left(\frac{1}{.60^2 + \frac{.81^2}{4 + .81^2}} - 1\right)^{-1} = 1.78. \qquad (19)$$

This implies that $p^+ = 1.78 \cdot 5 - 5 = 3.9$. When 4 items are added the resulting $\mathbf{P}^2$ will be about .83. So, a compensation for eliminating a mean-difference of about .80 is feasible.
The compensation of the elimination of a group mean-difference $d \geq .50$ by means of additional items $p^+ = kp - p$ is plotted for different levels of $\mathbf{P}^2 \geq .70$ and different levels of $\lambda_\xi$ in Figure 3. For $\lambda_\xi = .40$ a compensation is not possible with a few items (Figure 3 A), but for $\lambda_\xi = .60$ and $\lambda_\xi = .80$ the effect of eliminating moderate or large mean-differences on $\mathbf{P}^2$ can be compensated by a few additional items (Figure 3, B and C).

When a group mean-difference is eliminated and when it is impossible to compute a multiple-group factor analysis for the respective groups, it might be of interest to compute the reduced factor loading. Equation 13 can be transformed in order to correct the initial factor loadings, when the eliminated group mean-difference $d$ is known. This yields

$$\lambda_{\xi g}^2 - \frac{d^2}{4 + d^2} = \lambda_\xi^2. \qquad (20)$$

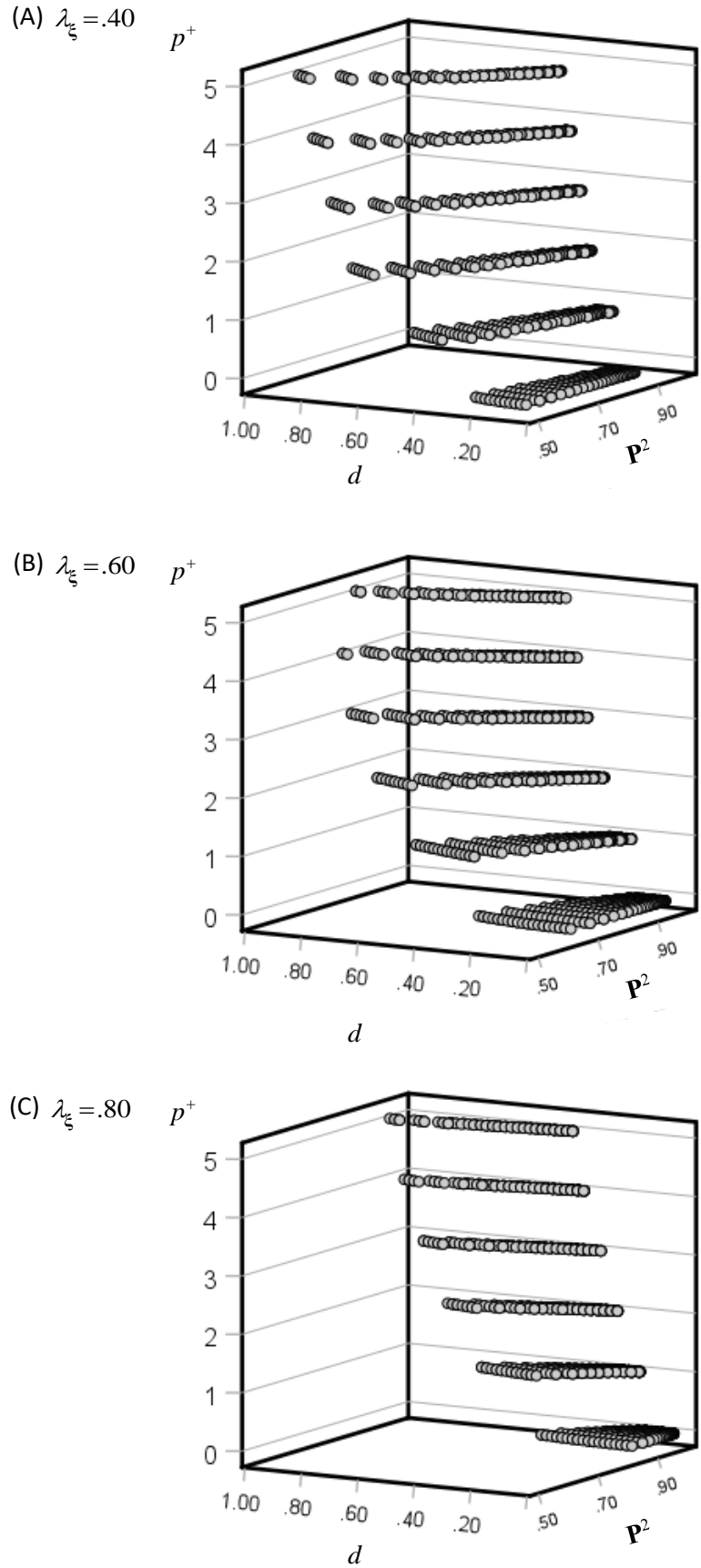

Figure 3. Number of additional items $p^+$ that is necessary maintain a given $\mathbf{P}^2 \geq .70$, when $d \geq .50$ is eliminated, (A) for $\lambda_\xi = .40$, (B) $\lambda_\xi = .60$, and (C) $\lambda_\xi = .80$.

## Discussion

The determinacy coefficient describes the common variance of a factor score predictor with the respective factor and is therefore an indicator of the validity of factor score predictors. The present study investigates to what degree the validity of factor score predictor may depend on a given mean-difference between two groups. This question is relevant when mean-differences between groups are eliminated by means of covariance analysis, regression, or group specific norms.

It is shown that in a one-factor model comprising the same group mean-difference on all observed variables, the common factor cannot be distinguished from a common factor representing the group mean-difference. This occurs even when the common factor and the group mean-difference have a zero covariance. Moreover, the effect of eliminating a group mean-difference $d$ on the common variance of the factor score predictor with the factor (determinacy coefficient) is investigated under the condition of equal common factor loadings. It is shown that for common factor loadings equal or larger than .60 the elimination of group mean-difference of a moderate effect size ($d = .50$) leads to only small decreases of the determinacy coefficient. It is, nevertheless, important to estimate the reduced factor score determinacy when group mean-differences in the factor score predictors are eliminated. In this context, an estimation of the reduced factor loadings can also be of interest.

Finally, a compensation-factor $k$ is proposed allowing for the estimation of the number of additional observed variables that are needed to recover the size of the determinacy coefficient before elimination of the group mean-difference. It turns out that for factor loadings equal or larger than .60 only a few additional items are needed in order to recover the initial determinacy coefficient after the elimination of moderate or large group mean-differences.

A limitation of the study is that it is only based on the best-linear (regression) factor score predictor (Krijnen et al., 1996). An investigation of the elimination of group mean-differences on the determinacy coefficients for other factor score predictors (Beauducel, & Hilger, 2022; Grice, 2001; Krijnen et al., 1996) may also be of interest. Other lines for further research are the investigation of the effect of unbalanced group-size, mean-differences between more than two groups, the effect of categorical observed variables, the effect of non-normal score distributions, and the effect of methods for the estimation of model parameters.

## Acknowledgement

This study was funded by the German Research Foundation (DFG), BE 2443/18-1.